\documentclass{elsart}

 \usepackage{graphicx}
 \usepackage{epsfig}

\usepackage{amssymb}

\begin{document}

\begin{frontmatter}

\title{
Ground State Baryons in $\tilde U(12)$ Scheme
}

\author{Muneyuki Ishida}
\address{Department of Physics, Meisei University, Hino, Tokyo 191-8506, Japan}

\begin{abstract}

The properties of ground-state baryons of light-quarks are investigated in
$\tilde U(12)$-classification scheme of hadrons, recently proposed by us.
In $\tilde U(12)$, in addition to the ordinary {\bf 56} of $SU(6)_{SF}$, 
the existence of the extra {\bf 56} with positive-parity and {\bf 70} with negative-parity
appear as ground-states in lower mass region.  
The $N(1440)$, $\Lambda (1600)$ and $\Sigma (1660)$ have the plausible properties
of masses and strong decay widths as the flavor-octet of extra {\bf 56},
while the ordinary radially-excited octet states are expected not to be observed as 
resonances because of their large predicted widths of the decays to extra {\bf 70} baryons.
The extra {\bf 70} baryons are not observed directly as resonances
for the same reason, except for only the $\Lambda (1406)$, of which 
properties are consistently explained through the singlet-octet mixing.
The baryon properties in lower mass regions are consistently explained in 
$\tilde U(12)$ scheme.

\end{abstract}

\end{frontmatter}


\section{Introduction}
\label{sec1}

   The spectroscopy of light-quark $qqq$ baryons is longstanding problem of hadron physics.
The non-relativistic quark model(NRQM) successfully explain the properties of
ground state {\bf 56}-multiplet of $SU(6)$ spin-flavor symmetry. 
On the other hand, NRQM predicts the negative-parity states as 
the next low-lying states from orbital excitation,
while the experiments show the clear evidence of the Roper resonance $N(1440)$
of the second nucleon state with positive-parity. 
The situation is similar for $\Delta$, $\Lambda$ and $\Sigma$ systems. 
The positive-parity $\Delta (1600)$, $\Lambda (1600)$ and $\Sigma (1660)$ have 
too light masses to be naturally assigned as radially excited states in NRQM. 
The mass of the negative parity $\Lambda (1405)$ is also too light to be assigned as 
the first excited {\bf 70}-multiplet in NRQM. 

   Recently\cite{[1]}, we have proposed a covariant level-classification scheme of hadrons
based on $\tilde U(12)$ group. In this scheme, the squared-mass spectra
of hadrons including light constituent quarks are classified as the representation of 
$\tilde U(12)$ spin-flavor group. 
In this scheme we have introduced the expansion bases of spinor wave functions(WF) of 
composite hadrons. Each spinor index corresponding to light quark freedom is expanded by 
free Dirac spinors $u_{r,s}(v_\mu )$.
\begin{eqnarray}
u_{+,s}(v_\mu ) &=& \left( \begin{array}{c} ch\theta \chi^{(s)}\\ 
                  sh\theta \mbox{\boldmath $n\cdot\sigma$}\chi^{(s)} \end{array} \right), \ \ \ 
u_{-,s}(v_\mu ) = \left( \begin{array}{c}  sh\theta \mbox{\boldmath $n\cdot\sigma$}\chi^{(s)}\\
                  ch\theta \chi^{(s)} \end{array} \right),
\label{eq1}
\end{eqnarray}
where $ch\theta , sh\theta =\sqrt{(\omega \pm 1)/2}$ and
the $v_\mu \equiv P_\mu /M=(\mbox{\boldmath $n$}\omega_3;i \omega )_\mu$ 
is the four velocity of the relevant hadron. 
This method is invented for keeping the Lorentz-covariance of the composite system.
(Concerning the $\tilde U(12)$ scheme and its group theoretical arguments,
 see our ref.\cite{[2]}.) 
We should note both $u_+$ and $u_-$ is necessary for expansion bases of
quark spinor index. They are called ur-citon spinors for historical reason\cite{ur}.

The $u_{+,s}$ corresponds to the ordinary spinor freedom appearing in NRQM, 
while the $u_{-,s}$ represents the relativistic effect. In the $\tilde U(12)$-classification scheme, 
the $u_{-,s}$ is supposed to appear as new degrees of freedom for light constituents,
being independent of $u_{+,s}$.
The freedom corresponding $r$ index of $u_{r,s}$ is called $\rho$-spin, 
while the ordinary Pauli-spin freedom described by $\chi^{(s)}$ is called $\sigma$-spin,
where $\rho \times \sigma$ corresponds to the $\rho\sigma$ decomposition of Dirac $\gamma$ matrices.
The index $s$ of $u_{r,s}$ represents the eigen value of $\sigma_3$, while the $r$ does 
the eigen value of $\rho_3$ at the hadron rest frame, where
$v_\mu =v_{0\mu}=(\mbox{\boldmath $0$};i)_\mu$.

Because of this extra $SU(2)$ spin freedom, the $SU(6)_{SF}$ is extended to $\tilde U(12)$,
or more precisely $U(12)_{\rm stat}$ at the hadron rest frame, as 
$U(12)_{\rm stat}\supset SU(2)_\rho\times SU(2)_\sigma\times SU(3)_F$
(see, ref.\cite{[2]}). 
The ground-state $qqq$ baryons and anti-baryons are assigned as the completely symmetric 
$({\bf 12}\times{\bf 12}\times{\bf 12})_{sym}={\bf 364}$ representation of $U(12)_{\rm stat}$ or
$\tilde U(12)$.
The corresponding flavor-spinor WFs $\Phi_{ABC}(v_\mu)$
are represented by the direct product of flavor and spinor WFs as
\begin{eqnarray}
\Phi_{ABC}(v) &\sim& |F\rangle_{abc} u_\alpha (v) u_\beta (v) u_\gamma (v),
\label{eq2}
\end{eqnarray}
where 
$A=(a,\alpha)$ etc. denote the (flavor,spinor) indices. 
$|F\rangle (u(v))$ represents the flavor(spinor) WF.
At the rest frame of hadron, the $u_\alpha (v)$'s are decomposed as $\rho$ and $\sigma$ spin WFs
denoted as $|\rho,\rho_3/2\rangle$ and $|\sigma_3/2\rangle$, respectively. 
The ground-state baryons in $\tilde U(12)$ are classified as 
${\bf 364}/2={\bf 56}_E+{\bf 70}_G+{\bf 56}_F$ in terms of $SU(6)_{SF}$.
The explicit forms of WFs are given in Table \ref{tab1}.

\begin{table}
\begin{center}
\begin{tabular}{clcc}
\hline
$SU(6)_{SF}$ & spin-flavor wave function & $B^{P\hspace{-0.28cm}\bigcirc}$ & $SU(3)_F$ \\ 
\hline
${\bf 56}_E$ & 
 $|\rho \rangle_S  |F\sigma\rangle_S=|\rho ,\frac{3}{2}\rangle_S |F\rangle_S|\sigma\rangle_S$ 
 & $\Delta_{3/2}^{+\hspace{-0.28cm}\bigcirc}$ & {\bf 10} \\
       & \ \ \ \ \ \ \ \ \ \ \ \ \ \ \ \ \ 
    $|\rho ,\frac{3}{2}\rangle_S \frac{1}{\sqrt{2}}(|F\rangle_\alpha |\sigma\rangle_\alpha +
     |F\rangle_\beta |\sigma\rangle_\beta )$ 
 & $N_{1/2}^{+\hspace{-0.28cm}\bigcirc}$ & {\bf 8} \\
\hline
${\bf 70}_G$ & 
$\frac{1}{\sqrt{2}}(|\rho\rangle_\alpha |F\sigma\rangle_\alpha 
       +|\rho\rangle_\beta |F\sigma\rangle_\beta )$ ;\ \ \  
    $|F\sigma \rangle_{\alpha (\beta )} = |F\rangle_S |\sigma\rangle_{\alpha (\beta )}$
 & $\Delta_{1/2}^{-\hspace{-0.28cm}\bigcirc}$ & {\bf 10} \\
  & \ \ \ \ \ \ \ \ \ \ \ \ \ \ \ \ \ \ \ \ \ 
\ \ \ \ \ \ \ \ \ \ \ \ \ \ \ \ \ \ \ \ \ \ \ \ \ \ \ \ \ \ \ \ \ \ 
    $|F\rangle_{\alpha (\beta )} |\sigma\rangle_S$ 
 & $N_{3/2}^{-\hspace{-0.28cm}\bigcirc}$ & {\bf 8}    \\
  & 
    $| F \sigma \rangle_{\alpha (\beta )} = \frac{1}{\sqrt{2}}
         ( \mp |F\rangle_\alpha |\sigma\rangle_{\alpha (\beta )} 
           +|F\rangle_\beta |\sigma\rangle_{\beta (\alpha )})$ 
 & $N_{1/2}^{-\hspace{-0.28cm}\bigcirc}$ & {\bf 8} \\
 & $|F\rangle_A  |\rho\sigma\rangle_A=|F\rangle_A
   \frac{1}{\sqrt{2}}(-|\rho ,\frac{1}{2}\rangle_\alpha |\sigma\rangle_\beta
   +|\rho ,\frac{1}{2}\rangle_\beta |\sigma\rangle_\alpha )$  
 & $\Lambda_{1/2}^{-\hspace{-0.28cm}\bigcirc}$ & {\bf 1} \\
\hline
${\bf 56}_F$  & 
$|\rho\rangle_S |F\sigma\rangle_S=|\rho ,-\frac{1}{2}\rangle_S |F\rangle_S|\sigma\rangle_S$ 
 & $\Delta_{3/2}^{+\hspace{-0.28cm}\bigcirc}$ & {\bf 10}   \\
  & \ \ \ \ \ \ \ \ \ \ \ \ \ \ \ \ \ 
$|\rho ,-\frac{1}{2}\rangle_S \frac{1}{\sqrt{2}}(|F\rangle_\alpha |\sigma\rangle_\alpha +
|F\rangle_\beta |\sigma\rangle_\beta )$ 
 & $N_{1/2}^{+\hspace{-0.28cm}\bigcirc}$  & {\bf 8} \\
\hline
\end{tabular}
\end{center}
\caption{Flavor-spinor WF of ground-state $qqq$-baryon.
The indices $S,\alpha (\beta),A$ represent the completely symmetric, 
partially symmetric(anti-symmetric) and completely anti-symmetric WFs of permutation group.}
\label{tab1}
\end{table}

The ${\bf 56}_E$ has all the three spinor indices with positive $\rho_3$ and $\rho$-spin WF
is given by the completely symmetric $|\rho ,\frac{3}{2}\rangle_S$.
This freedom corresponds to the ones appearing in NRQM, and the corresponding
states are called Pauli-states(or Paulons) . 
The ${\bf 56}_E$ includes the  $N(939)$-octet and $\Delta (1232)$-decouplet.
While ${\bf 70}_G({\bf 56}_F)$ have one index(two indices) with negative $\rho_3$ and are described
by the $\rho$-spin WFs $|\rho ,\frac{1}{2}\rangle_{\alpha (\beta)}$ ($|\rho ,-\frac{1}{2}\rangle_S$). 
These states are out of the conventional non-relativistic $SU(6)_{SF}$ framework, and are called
chiral states(or chiralons), since
the $u_{+ ,s}$ changes $u_{- ,s}$ by multiplying the chiral matrix $-\gamma_5=\rho_1$
as $\rho_1 u_{+ ,s}=u_{- ,s}$. 
We predict, in addition to the ordinary ${\bf 56}_E$, 
the existence of the extra ${\bf 56}_F$ with positive-parity and 
${\bf 70}_G$ with negative-parity in lower mass region since they are $S$-wave states.
There is an interesting possibility that all the problematic resonances mentioned above 
are explained as chiral states in $\tilde U(12)$ scheme.
In order to check this possibility,  
we consider the mass spectra and decay properties of ${\bf 56}_F$ and ${\bf 70}_G$.

\section{Mass Spectra}

In NRQM the mass spectra of ${\bf 56}_E$ multiplet are given by the Hamiltonian\cite{[3]}
\begin{eqnarray}
H_{NR} &=& \sum_{i=1}^3 m_i^C + \sum_{i<j} \frac{a}{m_im_j} 
           \mbox{\boldmath $\sigma^{(i)}\cdot\sigma^{(j)}$}.
\label{eq3}
\end{eqnarray}
%
We extend this $H_{NR}$ to the form applicable to the chiral $\rho$-spin space. 
Intuitively, we take the form,
\begin{eqnarray}
H &=& \sum_{i=1}^3 m_i^C + \sum_{i<j} \frac{(b_1+b_2\rho_3^{(i)})(b_1+b_2\rho_3^{(j)})a}{m_im_j}
      \mbox{\boldmath $\sigma^{(i)}\cdot\sigma^{(j)}$} + C_\chi .
\label{eq4}
\end{eqnarray}
The first two terms reduce to $H_{NR}$ in non-relativistic
$|\rho ,\frac{3}{2}\rangle_S=|\uparrow\uparrow\uparrow\rangle_\rho$ space
with constraint $b_1+b_2=1$.
The $C_\chi$ term is phenomenologically introduced to describe the overall mass splittings
between ${\bf 56}_F$ and ${\bf 56}_E$ and between ${\bf 70}_G$ and ${\bf 56}_E$.
The $C_\chi$ for ${\bf 56}_E$ is taken to be 0 as $C_E=0$, while two $C_\chi$'s, 
$C_F $for ${\bf 56}_F$ and $C_G$ for ${\bf 70}_G$, are taken as independent parameters. 

By taking the expectation value $\langle F\rho\sigma| H | F\rho\sigma \rangle$,
by WF given in Table \ref{tab1}, we obtain the mass formula of ground-state baryon systems. 
We use the seven parameters, which are determined as:
$m_n^C,m_s^C,A$ are from the masses of $N(939)$, $\Lambda (1116)$ and $\Delta (1232)$.
$r$ is from the mass of $\Omega (1672)$. $b_2,C_E$ are from the masses 
of $N(1440)$ and $\Sigma (1660)$. $C_G$ is from the mass of $\Lambda (1406)$. For $\Lambda (1406)$,
the singlet-octet mixing in ${\bf 70}_G$ coming from $T_{33}$ breaking of spin-spin interaction
is taken into account. 
We can predict the masses of all the other ground-state baryons in Table \ref{tab2}.

\begin{table}
\begin{center}
\begin{tabular}{l@{}l@{}c|l@{}l@{}c}
\hline
${\bf 56}_E$,${\bf 56}_F$ & $\langle H_{\rm spin}+C_\chi\rangle$ & $M_{\rm theor}$ 
          & ${\bf 70}_G$  & $\langle H_{\rm spin}+C_\chi\rangle$ & $M_{\rm theor}$ \\
\hline
$N(939)$         & $-3A$ & \underline{939} & 
   $\Delta_{1/2}$ & $-3A(1-\frac{8}{3}b_2)+C_G$ & 1284 \\ 
$\Lambda (1116)$ & $-3A$ &\underline{1116}& 
   $\Sigma^*_{1/2}$ & $-(2r+1)A(1-\frac{8}{3}b_2)+C_G$ & 1480 \\
$\Sigma (1195)$  & $-(4r-1)A$ & 1191 &  
   $\Xi^*_{1/2}$ & $-(2r+r^2)A(1-\frac{8}{3}b_2)+C_G$ & 1674 \\
$\Xi (1318)$ & $-(4r-r^2)A$ & 1338 &    
   $\Omega_{1/2}$ & $-3r^2A(1-\frac{8}{3}b_2)+C_G$ & 1863 \\
$\Delta (1232)$  & $\ \ 3A$ & \underline{1232} &   \\ 
$\Sigma^*(1385)$ & $\ \ (2r+1)A$ & 1371 & 
   $N_{3/2}$ & $\ \ 3A(1-\frac{4}{3}b_2)+C_G$ & 1474 \\
$\Xi^*(1533)$ & $\ \ (2r+r^2)A$ & 1518 &  
   $\Lambda_{3/2}$ & $\ \ (2r+1-2(r+1)b_2)A+C_G$ & 1620 \\
$\Omega (1672)$ & $\ \ 3r^2A$ & \underline{1672} & 
   $\Sigma_{3/2}$  & $\ \ (2r+1-\frac{10r+2}{3}b_2)A+C_G$ & 1624 \\
$N(1440)$    & $-3A\beta +C_F$ & \underline{1440} &  
   $\Xi_{3/2}$  & $\ \ (2r+r^2-\frac{10r+2r^2}{3}b_2)A+C_G$ & 1774 \\ 
$\Lambda (1600)$ & $-3A\beta +C_F$ & 1617 & \\
$\Sigma (1660)$  & $-(4r-1)A\beta +C_F$ & \underline{1660} &   
   $N_{1/2}$ & $-3A(1-\frac{4}{3}b_2)+C_G$ & 1249 \\
$\Xi       $ & $-(4r-r^2)A\beta +C_F$ & 1820 &  
   $\Lambda_{1/2}^{\bf 8}$ & $-(2r+1-4rb_2)A+C_G$ & $1494^*$ \\
$\Delta (1600)$  & $\ \ 3A\beta +C_F$ & 1608 &  
   $\Sigma_{1/2}$ & $-(2r+1-\frac{4r+8}{3}b_2)A+C_G$ & 1459 \\ 
$\Sigma^*$ & $\ \ (2r+1)A\beta +C_F$ & 1763 & 
   $\Xi_{1/2}$ & $-(2r+r^2-\frac{4r+8r^2}{3}b_2)A+C_G$ & 1652 \\ 
$\Xi^*$ & $\ \ (2r+r^2)A\beta +C_F$ & 1923 &    \\
$\Omega$ & $\ \ 3r^2A\beta +C_F$ & 2087 & 
   $\Lambda_{1/2}^{\bf 1}$ & $-(2r+1)A+C_G$ & $\underline{1406.5}^*$ \\
\hline
\end{tabular}
\end{center}
\caption{Mass formula and spectra of ground state baryons of light-quarks.
The sum of the $\sum_{i=1}^3m_i^C$ and $\langle H_{\rm spin}+C_\chi\rangle$ gives $M_{\rm theor}$.
The numbers with underlines are used as inputs to determine the values of parameters:
$m_n^C(m_s^C)=361.8(538.8)$MeV, $A(\equiv a/m_n^2)=48.8$MeV, 
$r(\equiv m_n/m_s)=0.6155$, $C_F(C_G)=438.4(275.8)$MeV
and $b_2=0.1757$. $\beta\equiv 1-4(2b_2-b_2^2)/3=0.5725$.
The values with $*$ include the effect of ${\bf 1}$ and ${\bf 8}$ mixing,
which is induced from the non-diagonal element of $H_{\rm spin}$, $-2A(1-r)(1-b_2)$.
The resulting $\Lambda (1406)$ WF is given by 
$|\Lambda (1406)\rangle = c_t |\Lambda_{1/2}^{\bf 1}\rangle +s_t |\Lambda_{1/2}^{\bf 8}\rangle$
with $(c_t,s_t)=(0.8133,0.5818)$.}
\label{tab2}
\end{table}

\section{Strong pionic and kaonic decays}

Next we consider the strong decays with one $\pi$ or $K$ emission.
The relevant decay amplitudes $T$ are obtained by using 
the low energy theorem of Nambu-Goldstone bosons as
$T=J_{A\mu}^i \frac{1}{f_{\phi^i}}\partial_\mu \phi^i$, where $J_{A_\mu}^i$ is the baryonic
axial-vector current, $\phi^i$ denotes the pseudoscalar octet, and $f_\phi$ are 
their decay constants. 
The $T$ is given by the quark effective interaction,
\begin{eqnarray}
{L}_{\rm eff}^{\rm quark} &=& -\bar q i\gamma_5\gamma_\mu \frac{\lambda^i}{2} q
                  \ \frac{1}{f_{\phi^i}} \partial_\mu \phi^i 
 \rightarrow   
  \bar u_q(p^\prime) i\gamma_5\gamma_\mu iq_\mu \frac{\lambda^i}{2f_{\phi^i}} u_q(p) \phi^i(q),
\label{eq5}\\
L_{\rm eff}^{\rm \prime\ quark} &=& -g_\pi \bar q (s-i\gamma_5\phi )  q
  \rightarrow g_\pi \bar u_q(p^\prime) i\gamma_5 \frac{\lambda^i}{\sqrt 2} u_q(p) \phi^i (q),
\ \ (\phi\equiv \frac{\phi^i\lambda^i}{\sqrt 2}),\ \ \ \ \ 
\label{eq6}
\end{eqnarray}
where we also give the momentum representation. 
The $L_{\rm eff}^{\rm quark}(L_{\rm eff}^{\rm \prime\ quark})$ is derivative
(non-derivative) type interaction. We may use either interaction or both, since
the $L_{\rm eff}^{\rm \prime\ quark}$ is reduced to $L_{\rm eff}^{\rm quark}$
by using Dirac equation of constituent quark spinor as
$g_\pi \bar u_q(p^\prime) i\gamma_5 u_q(p) 
 = -(g_\pi/2m_q) \bar u_q(p^\prime )i\gamma_5 \gamma_\mu iq_\mu u_q(p)$.
Thus, the $L_{\rm eff}^{\rm \prime\ quark}$ is equivalent to $L_{\rm eff}^{\rm quark}$. 
However, in $\tilde U(12)$-scheme, we cannot use the Dirac equation since there are
no constituent quark spinors. We have only urciton spinors with velocity of the hadron itself. 
Thus, in the effective interaction of urciton spinors, the non-derivative 
type coupling shows very different effects from the derivative type coupling.
We introduce both interaction with independent parameters,
\begin{eqnarray}
L_{\rm eff}^{urciton} &=& 
- g_{ND}/\sqrt 3\ \  \bar u_{r^\prime,s^\prime}(v^\prime_\mu)(s-i\gamma_5\phi )u_{r,s}(v_\mu)\nonumber\\
 && + g_D/\sqrt 3\ \  \bar u_{r^\prime,s^\prime}(v^\prime_\mu)(-iv^\prime\cdot\gamma )
           (s+i\gamma_5\phi )(-iv^\phi\cdot\gamma )u_{r,s}(v_\mu),
\label{eq7}
\end{eqnarray}
where $v^\phi_\mu \equiv q_\mu /m_\phi$ and the factor $1/\sqrt 3$ comes from the overlapping of 
color WFs. 
The $(-iv^\prime\cdot\gamma )$ factor takes the +1 for the Pauli states with $r^\prime =1$.
It is introduced for keeping the invariance of linear chiral transformation:
$u(v)\rightarrow exp\{ i \beta^i\lambda^i\gamma_5/2 \}u(v)$ and 
$(s+i\gamma_5\phi ) \rightarrow exp\{ i \beta^i\lambda^i\gamma_5/2 \}  (s-i\gamma_5\phi) 
 exp\{ i \beta^i\lambda^i\gamma_5/2 \} $.
The matrix elements between the spinors in Pauli states($r=+1$)
are expected to take the strength consistent with the low energy theorem, Eq.~(\ref{eq5}):
For pion emission, 
\begin{eqnarray} 
-\frac{m_\pi g_{ND}}{M+M^\prime}-g_D \simeq \frac{\sqrt 3 m_\pi}{\sqrt 2 f_\pi},
\label{eq8}
\end{eqnarray}
where $M(M^\prime)$ is the mass of initial(final) baryon.
We will check later this relation is satisfied approximately in our choice of parameters. 
For kaon emission we replace $g_D$ by $g_D\frac{m_K f_\pi}{m_\pi f_K}$. 

For the decays of flavor octet and singlet, we must consider the effect of $U_A(1)$ breaking
six-point interaction, $L_{\rm int}=-G_D\{ {\rm det}\ \bar q(1+\gamma_5)q
+{\rm det}\ \bar q(1-\gamma_5)q \}$.
By contracting the one $\bar q,q$ pair with the spinor WF of pseudoscalar meson,
we obtain the effective urciton interaction with the parameter $g_{det}$ 
\begin{eqnarray}
L_{\rm det}^{\rm urciton} &=& {ig_{det}}\  \epsilon_{a^\prime b^\prime c^\prime}
\epsilon^{abc} \phi_a{}^{a^\prime}\ 
  \bar u(v^\prime )^{b^\prime}u(v)_b\ \bar u(v^\prime )^{c^\prime}\gamma_5u(v)_c,
\label{eq9}
\end{eqnarray}
where $\epsilon^{abc}$ is the antisymmetric tensor of flavor $SU(3)$, which vanishes for
decouplet states. 

Next we consider the way to take the overlapping of spectator-quark indices. 
In the relativistic quantum mechanics, the overlapping leading to the probability amplitude 
is given by $u^\dagger ({\bf 0}) (\cdots ) u({\bf 0})$, 
where we have the hydrogen-like composite system in mind,
and {\bf 0} represents the velocity of the whole system. 
The system is non-covariant as a whole entity, and exactly speaking the overlapping is defined
in the frame of both initial and final hadrons being at rest. 

By using the four velocity $v_0=({\bf 0},i)$ at rest,
$u^\dagger ({\bf 0}) (\cdots ) u({\bf 0})=u^\dagger (v_0) (\cdots ) u(v_0)
 =\bar u (v_0)(-iv_0\cdot\gamma) (\cdots ) u(v_0)$, which is simply extended to the
case with the final hadron with velocity $v(\neq v_0)$ as
$\bar u (v)(-iv\cdot\gamma) (\cdots ) u(v_0)$.
We use this form for the spectator-quark transition where $(\cdots )=1$.

In the rest frame of initial baryon, the overlapping of WF is given by
\begin{eqnarray}
3 \langle F\rho\sigma (v_\mu )| &({\rm vertex})^{(1)}& 
   (-iv\cdot\gamma)^{(2)}(-iv\cdot\gamma)^{(3)} |F\rho\sigma (v_{0\mu})\rangle ,
\label{eq10}
\end{eqnarray}
where the interaction vertex, denoted as (vertex$)^{(1)}$, is given by 
Eqs.~(\ref{eq7}) and (\ref{eq9}).
We consider the total symmetricity of WF except for colors and the vertex is
inserted only to the first spinor-index with the factor 3. 
The formula of the decay widths between ${\bf 56}_{E}$ and ${\bf 56}_F$ are summarized 
in Table \ref{tab3}.
\begin{table}
\begin{center}
\begin{tabular}{l|ll}
 decay type  &  $\gamma$  &  $H$  \\
\hline
$D_E\rightarrow N_E \phi$ & $\frac{1}{3}|H|^2 f_{\alpha S}^2$ & 
     $g_{ND}+g_D (\omega^\pi + \frac{2M^\prime}{m_\pi}ch^2\theta)$ \\
$N_F\rightarrow N_E \phi$ & $\frac{2}{9}|H|^2$ & 
     $(g_{ND}-g_D \omega^\pi)f_\alpha +\frac{\sqrt 3}{2}g_{det}f_d$ \\
$N_F\rightarrow D_E \phi$  & $\frac{2}{9}|H|^2 f_{S\alpha}^2$ & 
     $-g_{ND}+g_D \omega^\pi$ \\
$D_F\rightarrow N_E \phi$ & $\frac{1}{9}|H|^2 f_{\alpha S}^2$ & 
     $-g_{ND}+g_D \omega^\pi $ \\
$D_F\rightarrow D_E \phi$ & $\frac{10}{9}|H|^2 f_{S}^2$ & 
     $g_{ND}-g_D \omega^\pi $    \\
\hline
\end{tabular}
\end{center}
\caption{Formula of $\gamma$ and decay amplitudes $H$, which lead 
the decay widths $\Gamma = \gamma\times 
         \frac{M^\prime \mbox{\boldmath $p^\prime$}}{\pi M}ch^4\theta sh^2\theta |H|^2$,
where we keep terms in the amplitude of the first order of $sh\theta$.
The $D_E(N_E)$ represents the decouplet(octet) in ${\bf 56}_E$.
Similarly $D_F(N_F)$ does those in ${\bf 56}_F$. 
The $f_i$'s are the factors from overlapping of flavor WF.
 }
\label{tab3}
\end{table}
In the actual calculation,
the urciton spinor of final baryon with $v_\mu$ is obtained  from the one with $v_{0\mu}$
by acting the Lorentz booster $\bar B(v)$ as
\begin{eqnarray}
\bar u(v)(-iv\cdot\gamma) = u^\dagger (v_0) \bar B(v), &\ &
   \bar u(v)=u^\dagger (v_0) \rho_3 \bar B(v),\ \ \ 
\label{eq11}
\end{eqnarray}
where $\bar B(v) = (ch\theta - \rho_1\mbox{\boldmath $n\cdot\sigma$}sh\theta )$. 
The $sh\theta$ appears in the amplitude for the process with $\rho_3$-flipping in $\bar B (v)$.
It is related with momentum of the final baryon through
$2 ch\theta\ sh\theta = \mbox{\boldmath $p^\prime$}/M^\prime$, which is suppressed by baryon mass
$M^\prime$, not by constituent quark mass $m_q$.
This suppression behavior is considered as a kind of conservation law in $\tilde U(12)$ scheme,
named $\rho_3$-line rule.

The vertex of pseudoscalar emission is proportional to $i\gamma_5=-i\rho_1$, which
is the $\rho_3$-flip interaction. Thus, the transitions within ${\bf 56}_E$(which 
has $|\rho,\uparrow\uparrow\uparrow\rangle_S$) and ${\bf 56}_F$(which has 
$|\rho,\uparrow\downarrow\downarrow\rangle_S$) 
requires at least one $\rho_1\mbox{\boldmath $n\cdot\sigma$}sh\theta$ factor in $\bar B(v)$. 
The relevant processes are $\rho_3$-violationg $P$-wave decays.

On the other hand, the decays of ${\bf 70}_G$(which has 
$|\rho,\uparrow\uparrow\downarrow\rangle_{\alpha (\beta)}$)
to $N_E\phi$ or $D_E\phi$ are $\rho_3$-conserving $S$-wave, and their decay amplitudes are 
proportional to $ch^3\theta$, not suppressed by $sh\theta$. 
Their decay widths $\Gamma$
generally becomes $\sim $a few GeV, which are quite large. 
The negative-parity ${\bf 70}_G$ baryons are expected not to be observed as resonances, but as backgrounds.
The decay of $\Lambda (1406)$ is the only exception. Its $\Sigma\pi$ decay width is given by
$\Gamma_{\Sigma\pi}=\frac{M^\prime \mbox{\boldmath $p^\prime$}}{4\pi M}H^2$ with
$H=-(c_t-s_t)(-g_{ND}+g_D\omega^\pi)-(c_t-\frac{s_t}{2})\frac{g_{det}}{\sqrt 3}$,
where $c_t$ and $s_t$ are the coefficients of singlet-octet mixing, given 
in the caption of Table \ref{tab2}.
Only $\Lambda (1406)$ has small width in ${\bf 70}_G$ because of its small decay phase space,
and partly because of the cancellations between $c_t$ and $s_t$ terms and 
between $g_{ND}$ and $g_{det}$ terms.


The parameters $g_{ND}$, $g_D$ and $g_{det}$ are determined from
the experimental decay widths of $\Delta (1232)\rightarrow N\pi$, 
$N(1440)\rightarrow N\pi$ and $\Lambda (1406)\rightarrow \Sigma\pi$, as
$(g_{ND},g_D,g_{det})=(-37.65,0.761,-22.39)$. We can see the relation 
Eq.~(\ref{eq8}) is approximately satisfied. Now we can predict the widths 
of all the other decay channels. The results are shown in Table \ref{tab4}. 
\begin{table}
\begin{center}
\begin{tabular}{l@{}ccc|l@{}c@{}ccc}
 decay mode  &  $f_{\alpha S}^2$  & $\Gamma_{\rm theor}$ & $\Gamma_{\rm exp}$  &
    decay mode  &  $f_\alpha$ & $f_d$ & $\Gamma_{\rm theor}$ & $\Gamma_{\rm exp}$ \\
\hline
$\Delta (1232)$$\rightarrow$$N\pi$ & $\frac{2}{3}$ & \underline{120} & $\sim$120 &
   $N(1440)$$\rightarrow$$N\pi$ & $-\frac{1}{\sqrt 6}$ & 0 & \underline{227.5} & $\sim$200 \\
$\Sigma^*(1385)$$\rightarrow$$\Lambda\pi$ & $\frac{1}{3}$ & 30 & 33 &
   $\Lambda (1617)$$\rightarrow$$\Sigma\pi$ & $-\frac{1}{\sqrt 2}$ & $\frac{1}{\sqrt 2}$ & 80 & 15$\sim$90\\
$\Sigma^*(1385)$$\rightarrow$$\Sigma\pi$ & $\frac{2}{9}$ & 3.5 & 4 &
   $\Lambda (1617)$$\rightarrow$$NK$ & 0 & $\frac{1}{\sqrt 3}$ & 70 & 20$\sim$45\\
$\Xi^*(1533)$$\rightarrow$$\Xi\pi$ & $\frac{1}{3}$ & 7.6 & 9.1$\pm$0.5 &
   $\Sigma (1660)$$\rightarrow$$\Lambda\pi$ & $-\frac{1}{\sqrt 6}$ & $\frac{1}{\sqrt 6}$ & 61 & \\
$N(1440)$$\rightarrow$$\Delta\pi$ & $\frac{4}{3}$ & 60 & 70$\sim$105 &
   $\Sigma (1660)$$\rightarrow$$\Sigma\pi$ & $\frac{1}{3}$ & $-1$ & 19 & \\
$\Lambda (1617)$$\rightarrow$$\Sigma^*\pi$ & 1 & 62 &  &  
   $\Sigma (1660)$$\rightarrow$$NK$ & $-\frac{2}{3}$ & 1 & 45 & 10$\sim$30\\
$\Sigma (1660)$$\rightarrow$$\Sigma^*\pi$ & $\frac{2}{9}$ & 27 &  &  
   $\Xi (1820)$$\rightarrow$$\Xi\pi$ & $-\sqrt{\frac{2}{3}}$ & $\sqrt{\frac{3}{2}}$ & 44 &  \\
$\Xi (1820)$$\rightarrow$$\Xi^*\pi$ & $\frac{1}{3}$ & 40 &  & 
   $\Xi (1820)$$\rightarrow$$\Lambda K$ & $\frac{1}{\sqrt 6}$ & $-\sqrt{\frac{2}{3}}$ & 0.3 &  \\
$\Lambda (1406)$$\rightarrow \Sigma\pi$ &       &   \underline{50} & 50$\pm$2 &
   $\Xi (1820)$$\rightarrow$$\Sigma K$ & $-\frac{1}{\sqrt 6}$ & 0 & 67 &  \\ 
\hline
\end{tabular}
\end{center}
\caption{Decay widths $\Gamma$ in MeV of the decays of $D_E$ and $N_F$ multiplet.
The decay of $\Lambda (1406)$ is also shown.
The values with underlines are used as inputs.
The experimental data are from ref.\cite{[4]}.
}
\label{tab4}
\end{table}

As can be seen in Table \ref{tab4}, the predicted values of decay widths seem to be consistent
with the present experimental data although they have large uncertainty.
We use $\Gamma (N(1440)\rightarrow N\pi )$ as input. However, its magnitude is predicted through
the pion low energy theorem and our assignment of $N(1440)$ as ground $N_F$ state.
If the $N(1440)$ is assigned as a radially excited $2S$ state $N(2S)$ as in the conventional
quark model\cite{Isg}, the decay width becomes much smaller than the $\Gamma_{\rm exp}$
because of the small overlapping factor of $1S$ and $2S$ WFs. 
The $\Lambda (1617)$ and $\Sigma (1660)$ also show the plausible properties as octet chiral states
in ${\bf 56}_F$. Their rather small widths compared with $N(1440)$
are explained by the cancellation due to determinant-type interaction.

On the other hand the total decay widths of $D_F$ multiplet are predited with the values
more than $\sim$500MeV, much larger than the other decay modes. Naively, in Table \ref{tab3}
the factor $\frac{10}{9}$ of $D_F\rightarrow D_E\phi$ decay is  an order of magnitude larger
than the other channels. This factor comes from the overlapping of $\sigma$-spin WFs.  
For instance, $\Delta (1608)$ have $\Delta \pi$ widths as $\sim$1.7GeV,
which is much larger than the width of the observed resonance $\Delta (1600)$.

We can successfully explain the property of $N(1440)$, which is assigned in $\tilde U(12)$ scheme
as a $S$-wave $N_F$ state out of the conventional $SU(6)_{SF}$ framework.
Our new assignment raises a question at the same moment, where the genuine radially-excited $N(2S)$
state exists. Experimentally there is no plausible candidate of $N(2S)$ observed below 2 GeV.

We can predict the mass of $N(2S)$ by using the covariant oscillator
quark model (COQM\cite{COQM}) with one-gluon-exchange potential\cite{Yamada}. 
The resulting $M_{N(2S)}$ is 
$\sim 1.74$GeV.\footnote{
In COQM the WFs are determined by the spin-independent 4-dimensional oscillator potential,
reproducing the linearly-rising Regge trajectories.\cite{COQM}
By using this WF, we can predict the magnitudes of Coulomb $V_C$, spin-spin $V_{SS}$,
orbit-orbit $V_{OO}$, tensor $V_T$ and spin-orbit $V_{SO}$ potentials.
The orbitally excited $L=2$ states are around $M_{1D}\sim 1.95$GeV, 
where $\Delta_{\frac{7}{2}^+}(1950)$ and $N_{\frac{5}{2}^+}(2000)$ are observed.
This $M_{1D}$ is used as input and by using the values of $V_C$, $V_{SS}$ and $V_{OO}$,
we obtain the $M_{N(2S)}$ as $1.74$GeV.
} 
By using this $M_{N(2S)}$ and the decay coupling constants determined in this section, 
the decay widths of $N(2S)$ can be predicted.
We obtain small $\Gamma_{N(2S)\rightarrow N\pi}\simeq 60$MeV 
because this is $\rho_3$-violating process and partly because of the small overlapping
factor from space-time WFs. The small $\Gamma$ is also
obtained in the other quark model\cite{Isg}.
However, in $\tilde U(12)$ scheme, we must consider the decays to the negative-parity 
${\bf 70}_G$ baryons, that is, $N(2S)\rightarrow N_{\frac{1}{2}}(1249)\pi$ and 
$\Delta_{\frac{1}{2}}(1284)\pi$ (See Table \ref{tab2}). 
These processes are $\rho_3$-conserving and
the large decay widths are predicted, with 360(290)MeV for the former(latter).
There is another decay mode such as $N(2S)\rightarrow N\sigma \rightarrow N\pi\pi$.
Totally the $N(2S)$ is predicted to have decay width $>0.7$GeV, and not to be observed as 
a resonance, but as a $N\pi\pi$ background. The other octet $2S$ baryons are also not expected 
to be observed for the same reason in our scheme. 
This prediction seems to be consistent with
the present experimental situation.

\section{Radiative decays of $\Delta (1232)$, $\Lambda (1406)$ and $N(1440)$}

Finally we consider the radiative decays of $\Delta (1232)$, $\Lambda (1406)$ and $N(1440)$,
observed experimentally.
The spin-current interaction is given in the covariant form by
\begin{eqnarray} 
L_{spin}^{\rm urciton} &=& \bar u(v)(-iv\cdot\gamma )A_\mu (q) 
     \frac{Q^{(i)}e}{2m_i}i\sigma_{\mu\nu}iq_\nu (b_1^\prime + b_2^\prime (-iv_0\cdot\gamma ) ) u(v_0).
\label{eq12}
\end{eqnarray}
This form aside from the factor $(b_1^\prime + b_2^\prime (-iv_0\cdot\gamma ) )$ is 
used in the analyses of 
radiative decays in COQM\cite{[5]} and in the analyses\cite{[2]} of $D_s$ system.
Analyses of magnetic moments of the nucleon give $m_n(m_s)=336(510)$MeV.
In order to keep this result we require a constraint 
$b_1^\prime + b_2^\prime = 1$.
As a result,
the interaction vertex between $u^\dagger ({\bf 0})$ and $u({\bf 0})$ is given by\\
$\frac{e |\mbox{\boldmath $q$}|}{\sqrt 2 m_n}
\bar B(v) Q_M \sigma_\mp (1\pm \rho_1) (b_1^\prime + b_2^\prime \rho_3 )$,
where the sign corresponds to the helicity $\mp 1$ of the photon emitted to the 
$-z$ direction. There is no effect of $b_2^\prime$ term for the transition between Pauli-states,
while it gives sizable contribution for the decays of chiral states with $\rho_3$-flipping.

The parameter $b_2^\prime$ is taken to be $b_2^\prime =0.136$, which is determined by the width of
$\Lambda (1406)\rightarrow \Lambda \gamma$ as input. Now we can predict 
the radiaitve decay amplitudes $A_{\frac{1}{2}}$ or $A_{\frac{3}{2}}$,
where $A$ is related with the ordinary Lorentz-invariant amplitude $T$ as 
$A=T/\sqrt{2|\mbox{\boldmath $q$}|}$. We show only the numerical results in Table \ref{tab5}.

\begin{table}
\begin{center}
\begin{tabular}{lcc|cc}
Process  & $A_h^{\rm theor}$ & $A_h^{\rm exp}$ & 
        $\Gamma^{\rm theor}$ & $\Gamma^{\rm exp}$ \\
\hline
$A_{\frac{3}{2}}(\Delta (1232)\rightarrow N\gamma )$  & 
  -0.219 & -0.255$\pm$0.008 & 516 & 624$\sim$720 \\
$A_{\frac{1}{2}}(\Delta (1232)\rightarrow N\gamma )$ & 
  -0.126 & -0.135$\pm$0.006 &     &  \\
\hline
$A_{\frac{1}{2}}(\Lambda (1406)\rightarrow \Lambda\gamma )$ & 
   \underline{0.040} & 0.040$\pm$0.005 & \underline{27} & 27$\pm$8  \\
$A_{\frac{1}{2}}(\Lambda (1406)\rightarrow \Sigma^0\gamma )$ & 
   0.057 &  0.031$\pm$0.005  &  33  &  10$\pm$4  \\
 &       &  0.047$\pm$0.006  &      &  23$\pm$7  \\
\hline
\end{tabular}
\end{center}
\caption{The radiative decay amplitudes $A_h$  and 
decay widths $\Gamma$ for 
$\Delta (1232)\rightarrow N\gamma$ and $\Lambda (1406)\rightarrow \Lambda\gamma ,\Sigma^0\gamma$. 
$A_h(\Gamma)$ is given in the unit of GeV$^{-\frac{1}{2}}$(keV).
$\Gamma_\gamma = \frac{\mbox{\boldmath $q$}^2}{\pi}
\frac{1}{2J_I+1}\frac{2M^\prime}{M}( |A_{\frac{1}{2}}|^2+|A_{\frac{3}{2}}|^2 )$.
The experimental data of $\Delta(\Lambda (1406))$ decay is from ref.\cite{[4]}(\cite{[6]}).
There is two-fold ambiguity of amplitude of $\Lambda (1406)\rightarrow \Sigma^0\gamma$. }
\label{tab5}
\end{table}

It is remarkable that the amplitude of $\Delta (1232)\rightarrow N \gamma$ 
is greatly improved from the prediction by NRQM, where
$A_{\frac{3}{2}}=-\sqrt{|\mbox{\boldmath $q$}|/6}\ e/m_n =0.187$GeV$^{-\frac{1}{2}}$
leading $\Gamma =$377keV, which is much smaller than the experimental value.
In our covariant $\tilde U(12)$ scheme we obtain the relativistic recoil factor
$ch^3\theta+ch^2\theta sh\theta =1.17$ for the amplitude, leading to the great improvement, 
$\Gamma =516$keV. 
There is no effect from $b_2^\prime$ for $\Delta$ decay, while
there are sizable cancellation between $b_1^\prime$ and $b_2^\prime$ terms 
for $\Lambda (1406)$ decays. 
If we fix $b_2^\prime =0$ we obtain $\Lambda \gamma (\Sigma^0\gamma)$ width 
as 52(64)keV, which are not so much different from the experimental values.

On the other hand, for $N(1440)\rightarrow N\gamma$ decay,
the amplitude vanishes until the order of 
$ch^2\theta sh\theta$. There is a contribution of order $ch\theta sh^2\theta$,
however, because of the suppression by $sh\theta$, the predicted $A_{\frac{1}{2}}$ is 
of an order smaller than the experimental value.
This problem may be possibly resolved by considering the small mixing of 
$N_E$ WF to the $N_F$ WF in $N(1440)$, of which effect is proportional to $ch^3\theta$.
We consider in this paper only ideal case
that $N(939)$($N(1440)$) is purely $N_E$ Pauli state($N_F$ chiral state). 
This problem will be considered elsewhere.         

\section{Concluding Remarks}

In this letter we study the spectra and decay properties of 
light-quark baryon systems in $\tilde U(12)$ classification scheme of hadrons. 
The $N(1440)$, $\Lambda (1600)$ and $\Sigma (1660)$
have the property of the octet chiral states in the extra ground-state ${\bf 56}_F$ multiplet.   
The existence of $\Xi (1820)$ in the same multiplet is predicted 
with the decay width 40 MeV to $\Xi^* \pi$ channel,  and
the negligible decay width to $\Lambda K$ channel. 

The negative parity ${\bf 70}_G$ multiplet has generally very wide widths, 
and are expected to be not observed as resonances, except for $\Lambda (1406)$,
where the cancellation due to $U_A(1)$ breaking interaction diminishes 
its decay width to $\Sigma\pi$.
These results suggest the $\tilde U(12)$ representation is actually realized in
baryon spectroscopy of light quarks.
Especially the parity-inversion problem of excited nucleon spectra is naturally
resolved in $\tilde U(12)$ scheme, since
the mass and strong decay property of $N(1440)$ are explained as
the chiral $S$-wave $N_F$ state.
Ordinary radially excited $N(2S)$ state is not observed because of its large predicted widths
of the decays to chiral ${\bf 70}_G$ baryons.     

The property of $\Delta (1600)$ cannot be explained as the ground-state chiralon.
However, we expect the existence of $P$-wave chiral state in this energy region,
and the width of $\Delta (1600)$ may be explained from the possible mixing effect
to this $P$-wave chiral state.
This problem will be considered elsewhere.

{\it The author would like to express his sincere gratitude to Professor S. Ishida and 
Professor K. Yamada for their useful comments.}

\end{document}